\documentclass{aastex63}

\usepackage{ulem}

\received{\today}
\revised{\today}
\accepted{}
\submitjournal{Publications of the Astronomical Society of the Pacific}

\shorttitle{GReX}
\shortauthors{Connor et al.}
\graphicspath{{./}{figures/}}

\begin{document}

\title{Galactic Radio Explorer: an all-sky monitor for bright 
radio bursts}

\correspondingauthor{Liam Connor}
\email{liam.dean.connor@gmail.com}

\author[0000-0002-7587-6352]{Liam Connor}
\affiliation{Cahill Center for Astronomy and Astrophysics, MC 249-17, California Institute of Technology, Pasadena CA 91125, USA}

\author[0000-0003-4652-7038]{Kiran A. Shila}
\affiliation{Cahill Center for Astronomy and Astrophysics, MC 249-17, California Institute of Technology, Pasadena CA 91125, USA}

\author[0000-0001-5390-8563]
{Shrinivas R. Kulkarni}
\affiliation{Cahill Center for Astronomy and Astrophysics, MC 249-17, California Institute of Technology, Pasadena CA 91125, USA}

\author[0000-0002-0204-2891]{Jonas Flygare}
\affiliation{Onsala Space Observatory, Department of Space, Earth and Environment, Chalmers University of Technology \\ SE-41296 Gothenburg, Sweden}

\author{Gregg Hallinan}
\affiliation{Cahill Center for Astronomy and Astrophysics, MC 249-17, California Institute of Technology, Pasadena CA 91125, USA}
\affiliation{Owens Valley Radio Observatory, MC 249-17, California Institute of Technology, Pasadena CA 91125, USA}

\author{Dongzi Li}
\affiliation{Cahill Center for Astronomy and Astrophysics, MC 249-17, California Institute of Technology, Pasadena CA 91125, USA}

\author{Wenbin Lu}
\affiliation{Cahill Center for Astronomy and Astrophysics, MC 249-17, California Institute of Technology, Pasadena CA 91125, USA}
 
\author{Vikram Ravi}
\affiliation{Cahill Center for Astronomy and Astrophysics, MC 249-17, California Institute of Technology, Pasadena CA 91125, USA}
 
\author{Sander Weinreb}
\affiliation{Cahill Center for Astronomy and Astrophysics, MC 249-17, California Institute of Technology, Pasadena CA 91125, USA}

\begin{abstract}
We present the Galactic Radio Explorer (GReX), an all-sky monitor to probe the brightest bursts in the radio sky. Building on the success of STARE2, we will search for fast radio bursts (FRBs) emitted from Galactic magnetars as well as bursts from nearby galaxies. GReX will search down to $\sim$\,ten microseconds time resolution, allowing us to find new super giant radio pulses from Milky Way pulsars and study their broadband emission. The proposed instrument will employ ultra-wide band (0.7--2\,GHz) feeds coupled to a high performance (receiver temperature $<10\,$K) low noise amplifier (LNA) originally developed for the DSA-110 and DSA-2000 projects. In GReX Phase I (GReX-I), unit systems will be deployed at Owens Valley Radio Observatory (OVRO) and Big Smoky Valley, Nevada. Phase II will expand the array, placing feeds in India, Australia, and elsewhere in order to build up to continuous coverage of nearly 4$\pi$ steradians and to increase our exposure to the Galactic plane. We model the local magnetar population to forecast for GReX, finding the improved sensitivity and increased exposure to the Galactic plane could lead to dozens of FRB-like bursts per year.

\end{abstract}


\section{Introduction} \label{sec:intro}

The advent of wide-field, broad band radio 
surveys combined with our ability to search data at 
high time resolution has led to a number of novel discoveries. The fast 
radio burst (FRB) phenomenon in particular has radically changed the 
radio astronomy landscape \citep{cc19, petroff-2019}. In response, many FRB surveys 
have been built or proposed.

The detection of a Galactic FRB from SGR\,1935+2154 
by both STARE2 and the Canadian Hydrogen Intensity Mapping Experiment (CHIME/FRB) 
was the most significant step to date in connecting 
extragalactic FRBs to a known phenomenon \citep{CHIME_SGR1935, brb+20}. The value of 
having such objects nearby is difficult to overstate, and detecting more ultra-bright 
bursts is essential to understanding the connection between magnetars and FRBs. 

We propose the Galactic Radio Explorer (GReX, rhymes with ``T-Rex'') as a complement to deeper, high-spatial resolution 
surveys. This is similar to how X-ray all-sky monitors and more sensitive instruments work symbiotically; GReX will detect rare ultra-bright Galactic bursts that 
cannot be discovered without nearly continuous all-sky monitoring. We are therefore building on the STARE2 design \citep{bochenek-2019}, but with greater sensitivity, a
five-times larger bandwidth, and clusters of antennas dispersed around the world. 
The GReX design is meant to maximize the detection rate of $\sim$\,MJy bursts per unit cost.

\begin{figure}[htbp] 
 \centering
  \includegraphics[width=4.33in]{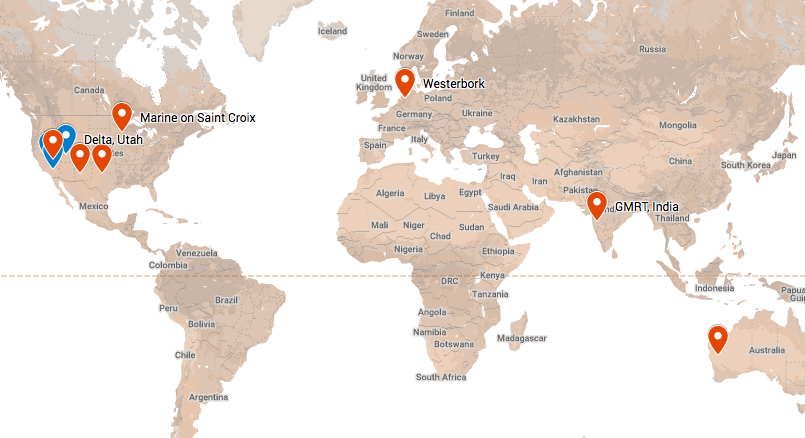}\qquad\\
  \includegraphics[trim=0 0 0 0in,width=5in]{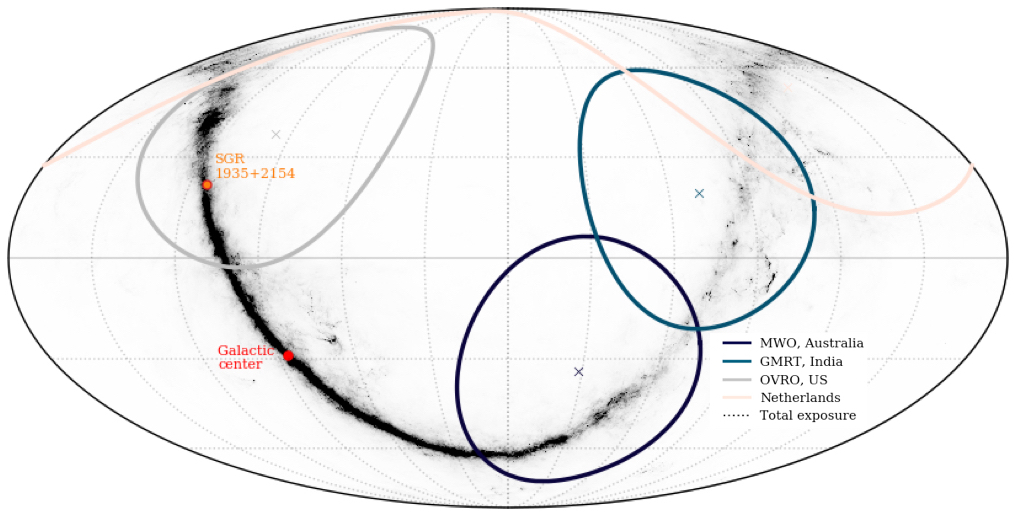}
   \caption{\small The layout of antenna clusters for GReX-I (blue) and GReX-II (red) is shown in the top figure. The bottom figure shows the instantaneous primary beam coverage of GReX clusters around the world, plotted over the Planck 857\,GHz map (as a proxy for molecular gas and thus star-formation and Galactic magnetars). The first Galactic FRB
   (SGR\,1935+2154) is marked in orange color. A hypothetical configuration of GReX-II, with three stations in India (marked ``GMRT"),
   Australia, and the Netherlands each, would nicely cover the inner
   Galaxy (where bulk of the magnetars are located). It would also increase the coverage of the Northern sky.}
 \label{fig:GRE_Exposure}
\end{figure}

FRBs have been detected over the frequency range 0.1\,GHz to 8\,GHz, 
though not simultaneously \citep{gajjar-2018, lofar-apertif, lofar-chime}.
The angular distribution is approximately isotropic and the daily all-sky rate
is $\sim$\,$10^3$ with fluence above a few Jy\,ms \citep{petroff-2019}. The typical observed pulse
widths are a few milliseconds, set by the $\sim$\,ms back-ends of most blind surveys; some FRBs are known to be tens to hundreds of 
microseconds in duration. A subset of 22 FRBs is known to repeat\footnote{https://www.chime-frb.ca/repeaters}\citep{fonseca-2019,chime-repeater-2019}, 
and two of those repeaters appear to 
do so periodically on weeks to months timescales \citep{CHIME_periodic, rajwade-2020}. It is still unclear if repeating FRBs and those that have not been seen to repeat form two physically distinct classes, though there is 
evidence that their pulse morphology \citep{pleunis-2020} and
widths have different statistical distributions \citep{fonseca-2019, connor-2020}.
Whether all FRBs are repeaters but with a large variation
in time between bursts, or if there is a population of genuine once-off FRBs, 
remains one of the main outstanding questions in the field.

\begin{figure}[htbp] 
\centering
   \qquad
 \includegraphics[width=0.25\textwidth]{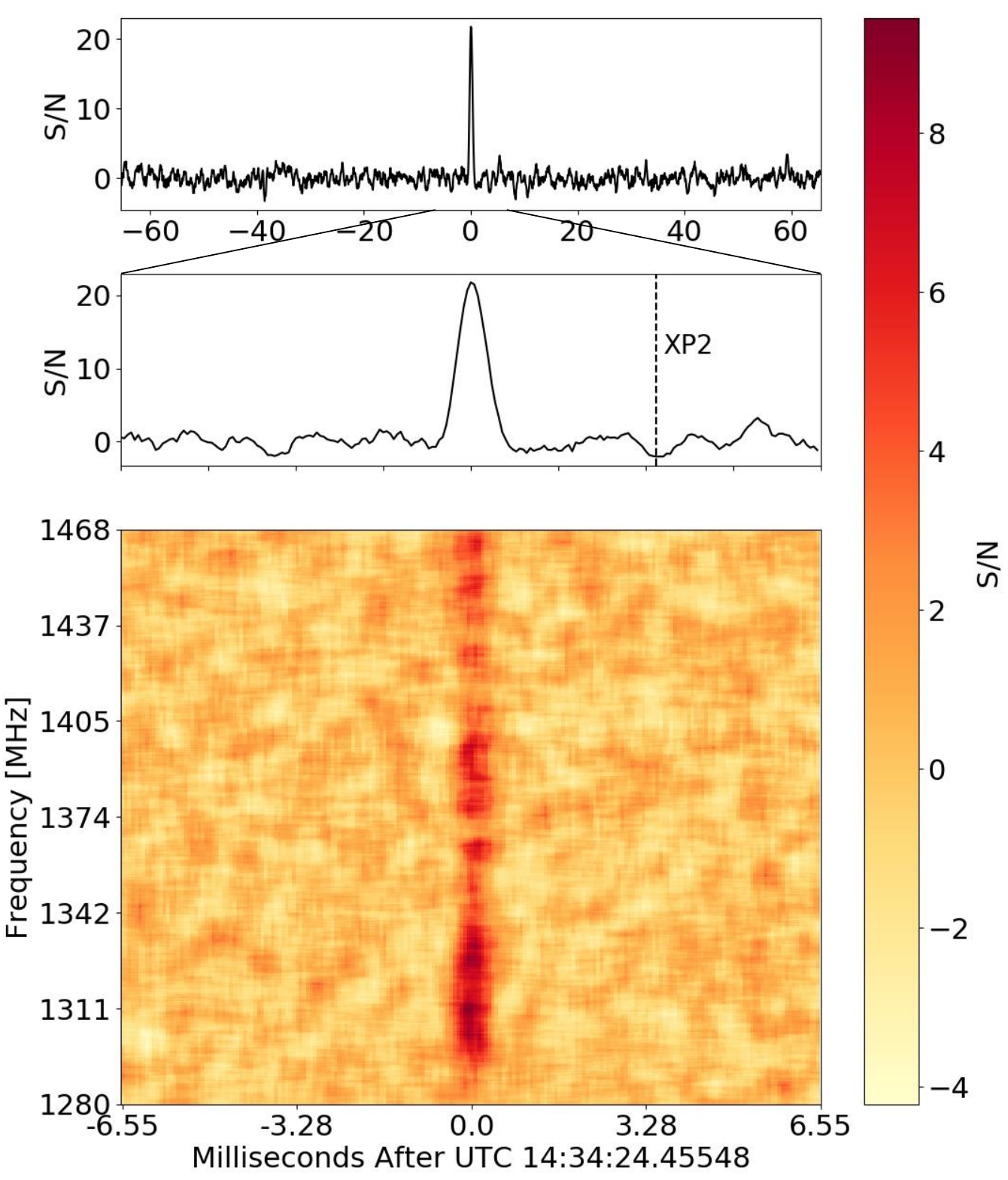}
 \includegraphics[width=0.45\textwidth]{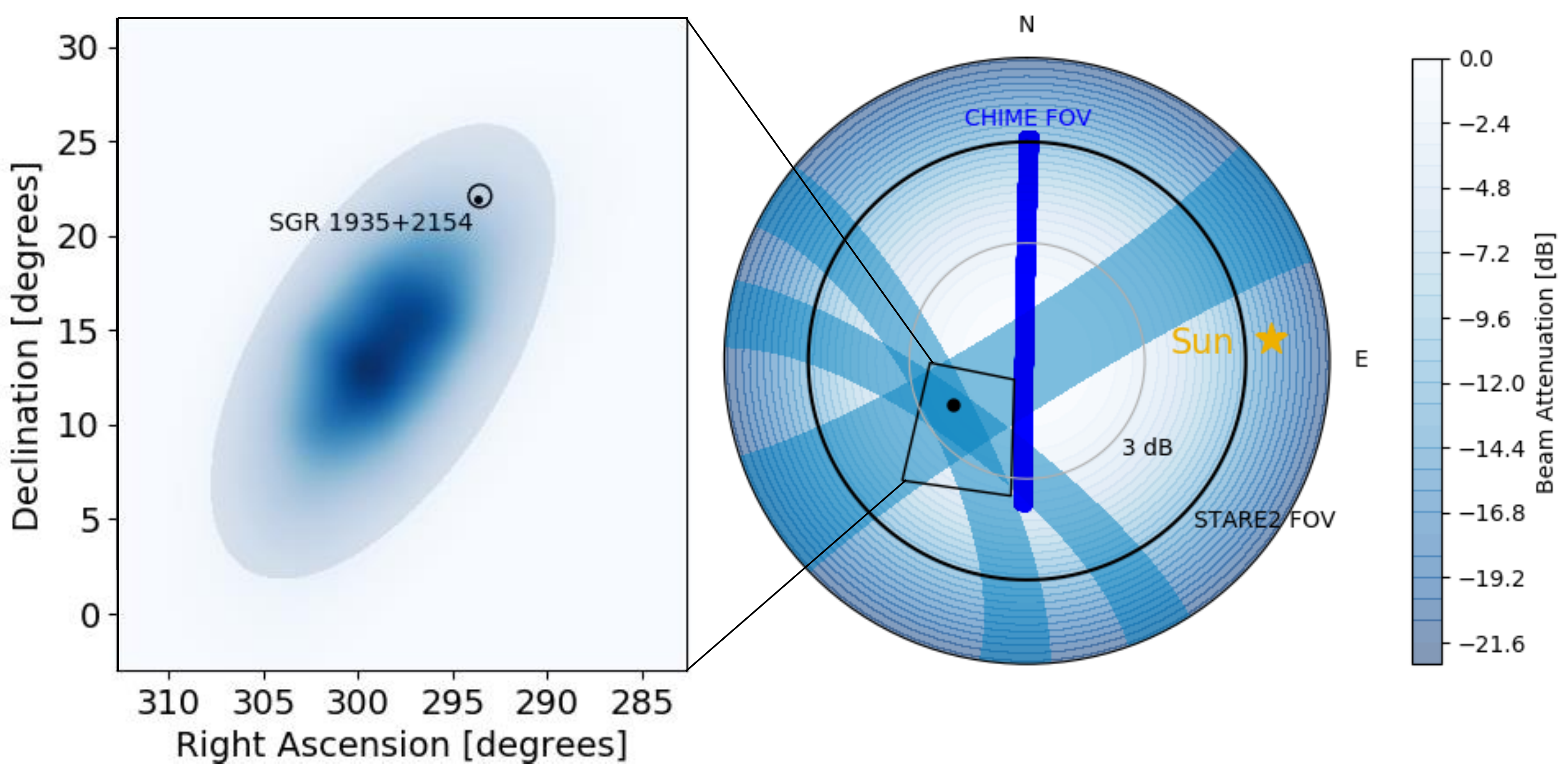}
   \caption{\small The dynamic spectrum of FRB\,200428 from STARE2 (left), along with the combined 
   localization region from its and CHIME/FRB's detection (right). The STARE2 burst is centered on the
   X-ray emission.}
 \label{fig:SGR1935+2154_CHIME_STARE2}
\end{figure}

The events leading up to and during April 28, 2020 resulted in a dramatic
link between magnetars and FRBs.  
On 27 April 2020, the Swift Burst
Alert Telescope reported multiple bursts from the soft $\gamma$-ray
repeater (SGR) 1935+2154, signaling that the magnetar had entered
a phase of heightened activity. The next day the CHIME/FRB collaboration
reported a dispersed burst with $\mathcal{F}$ of few kJy\,ms
(0.4--0.8\,GHz) in a side-lobe announced via ATel \citep{scholz-2020}, but this 
fluence value was later revised by a factor of $10^2$ \citep{CHIME_SGR1935}. 
The daily inspection of recorded STARE2 triggers
was then expedited, and ST\,200428A was found 
at approximately at the same time and dispersion measure (DM) as the
CHIME/FRB event. However, ST\,200428A had a {\it fluence that was a
one thousand times greater} than that reported by CHIME/FRB
(Figure~\ref{fig:SGR1935+2154_CHIME_STARE2}). Because this event 
was found in the complex sidelobes of CHIME, 
its fluence was simpler to measure in the STARE2 data. 
On 30 April 2020, the Five hundred metre Aperture Spherical Telescope
(FAST) reported a weak (0.06 Jy ms) radio pulse and localized to
SGR 1935+2154, with a DM consistent with the CHIME/FRB and STARE2 events.

Shortly thereafter, a constellation of space-borne instruments
reported a one-second-long X-ray (1--250\,keV) burst from the
direction of SGR 1935+2154 that occurred at precisely the same time
as the CHIME/FRB bursts and ST 200428A. This partnership between
radio facilities \citep{CHIME_SGR1935, brb+20, zjm+20}
is summarized graphically in Figure~\ref{fig:SGR1935+2154_CHIME_STARE2}.
Following the mega Janksy burst of 28 April, SGR\,1935+2154 emitted
intermittently bursts with fluence of 100\,Jy\,ms \citep{ksg+20} 
and has now finally become an intermittent pulsar at 
the X-ray period of 3.25\,s and fluence of 40\,mJy\,ms
\citep{zwz+20}. Clearly, magnetars are capable
of emitting mega Jansky bursts but not all magnetar X-ray flares are
accompanied by intense radio bursts. We need more observations of
this phenomenon to understand the currently murky X-ray-radio
connection.

In this paper we first discuss the science that can 
be done with a global network of antennas 
searching the sky continuously at tens of microseconds. 
We then describe the novel hardware and digital 
back-end that have been developed for DSA-110 and DSA-2000, which 
will be modified for GReX. This includes an ultra-wide band feed, 
extremely low-noise amplifiers (LNAs), and a sub-band single pulse search 
strategy. In Sect.~\ref{sect-forecast},
we model the Galactic pulsar and magnetar distribution in order 
the forecast FRB and giant radio pulse science that will emerge from 
GReX.

\section{Science goals}
\label{sect-science}
X-ray astronomy has shallow but very wide-field ``all-sky monitors"
or ASMs (e.g., Rossi X-ray Timing Explorer; MAXI) and also highly
sensitive instruments (e.g., Rossi Proportional Counting Array;
Chandra X-ray Observatory). The ASMs have historically played a
major role in identifying rare but bright events that are missed
by narrower field instruments.  In the same way, as demonstrated
by STARE2, radio astronomy would benefit from having a powerful
radio ASM. GReX will fill this role.

Now let us examine the science drivers for GReX in some
detail. \citet{connor-2019} has analyzed the performance of
FRB studies and argues that there likely exist many temporally narrow bursts missed
due to instrumental smearing. Giant radio
pulses (GRP) from pulsars have typical widths of microseconds but a few at the nanoseconds timescale have been detected \citep{spb+04}. The computational
complexity is enormous for major FRB surveys like those of ASKAP and CHIME/FRB, 
which search many radio beams and seek both to localize FRBs and detect them at a high rate. 
As a result, these searches limit their blind detection temporal resolution for detection to
$\sim$\,1\,ms. In contrast, each unit of GReX receives
only one signal stream with modest daily upload of candidates to
Caltech. Thus, each GReX unit can undertake
sub-millisecond pulses and in due course routinely explore the radio
sky at greater resolution. Next, FRBs appear to have
strong frequency structure. The usual approach of realizing higher
sensitivity as $\sqrt{B}$ where $B$ is the bandwidth is no longer
optimal. However, by undertaking sub-band searches, the ultra-wideband 
of GReX acts as a very broad frequency field of view. 

\vspace{0.25cm}
\noindent{\bf Galactic Magnetars}
The mega Janksy burst ST\,200428A 
from SGR\,1935+2154 solidified a bridge between the enigmatic 
extragalactic FRB phenomenon and a known physical object.
It also fills in an important 
chasm in the luminosity function of coherent radio pulses, as 
it was considerably more energetic than any known Galactic burst, 
but a couple of orders of magnitude less energetic than 
the weakest known FRBs. As demonstrated by the multi-telescope, multi-wavelength campaign 
on SGR\,1935+2154, the value of having such a magnetar nearby 
cannot be overstated. Even the highly-active repeating FRB\,180916.J0158+65, at just $\sim$\,150\,Mpc, is much too far to reasonably expect a high-energy detection. The recent discovery of the low-DM repeater FRB\,20200120E came 
within $\sim$\,20\,kpc of M81 \citep{m81}. The source was localized 
to a globular cluster in M81 \citep{kirsten2021} at 
3.6\,Mpc, falling roughly halfway between SGR\,1935+2154 and the FRB\,180916.J0158+65 in logarithmic distance. Still, it is not yet known if Galactic FRBs are physically identical to extragalactic FRBs, rather than just phenomenologically similar. Therefore, it is essential that we continue to 
capture Galactic FRBs and dig deeper into their luminosity function 
with the sensitivity improvements of GReX.

The dynamic spectra of extragalactic FRBs show several distinct features.
They are often band-limited, with downward drifting subpulses \citep{hessels-2018}, known as the ``sad trombone'' effect. 
While CHIME/FRB's detection of FRB\,200428 had significant time and frequency 
structure, it is not currently known if the mega bursts emitted by 
Galactic magnetars have similar dynamic spectra to extragalactic FRBs.
The ultra-wideband receivers of GReX 
allow us to ``catch'' the narrowband bursts that 
STARE2 might have missed, but also to study their dynamic spectra over a 
3:1 band---five times larger than the band of STARE2---and with almost 100 times better 
temporal resolution.

If Galactic mega bursts such as ST\,200428A are found to 
be the same physical phenomenon as extragalactic FRBs, we will be able 
to answer major open questions in the FRB field. For example, is the coherent radio emission produced in the neutron star's
magnetosphere, or is it produced in a relativistic shock well outside of the light cylinder? What is the origin of periodic activity in repeating FRBs? If bright bursts from 
Galactic magnetars are found to be meaningfully distinct 
from other FRBs, then that is proof that multiple 
mechanisms can produce $\sim$\,10$^{30}$\,erg\,Hz$^{-1}$ radio pulses; this would be evidence for the multiple-class 
interpretation of the FRB population.

\vspace{0.25cm}
\noindent{\bf Super-giant pulses from Galactic pulsars} Young
pulsars like the Crab and millisecond pulsars (MSPs) like PSR\,1937+214
are known to emit giant radio pulses.  Let us define ``super-giant
pulses'' (SGP) as those with fluence higher than $1\rm\, MJy\, \mu
s$. So far, only the Crab pulsar is known to emit SGPs at the GReX 
observing frequencies. The Crab GRP
rate is a power-law function of $\mathcal{F}$;
from \cite{bc19}, we expect $N(>\mathcal{F})
= 10^{-2}\,{\rm hr^{-1}} (\mathcal{F}/{10\, \rm MJy\,\mu s})^{-1.8}$. 
Assuming the typical pulse is broad band 
in frequency and that known sources such as the Crab will be coherently dedispersed, GReX could 
detect a 1\,$\mu$s pulse at a fluence of a few MJy$\,\mu$s. From the Crab we
expect a rate of,

\begin{equation}
\mathcal{R}_{\mathrm{det}}(>\mathcal{F}) \approx 10^2\,{\rm yr}^{-1}\, (X/5)^{1.8},
\end{equation}

\noindent where $X$ is the sensitivity improvement of GReX over STARE2.
Thus GReX should detect some of the brightest super-giant pulses from Crab-like
pulsars in our Galaxy, as well as the LMC and SMC. We note that the Crab's DM varies by $\sim$\,$10^{-2}$\,pc\,cm$^{-3}$ on timescales 
of roughly 1 year \citep{kuzmin2008}, leading to a dispersion delay 
across our band of $\sim\,7\mu$s and a reduction of S/N during coherent 
dedispersion. To account for this effect, GReX will use an up-to-date 
DM for its coherently dedispersed sources, 
using higher-sensitivity instruments such as DSA-110 to monitor 
DM variation.
It is also salient that some GRP-emitting pulsars 
exhibit a ``kink'' in their energy distribution, such 
that the powerlaw $N(>\mathcal{F}_{\rm th})$ flattens in the 
high-energy tail, making ultra-bright bursts more 
common \citep{Mahajan-2018}. 

GRPs have been detected from at least ten 
sources, mostly either young pulsars or MSPs \citep{kuzmin2007, Mahajan-2018, kuiack-0950}. \citet{Staelin-1968} discovered the Crab pulsar via 
its GRPs only days before an independent group at Arecibo found 
its normal periodic emission in a 
custom integer FFT developed for pulsar searching. This history was 
recounted by \citet{lovelace}. All GRPs 
were detected \textit{after} the pulsar or its supernova 
remnant were discovered, 
sometimes in targeted searches of pulsars with large 
light-cylinder magnetic field strengths \citep{knight-2005}. 
Pulsars are typically not discovered blindly via their 
giant pulses.
The Crab is $10^3$ years old, and we expect $\mathcal{O}(10)$ such young neutron stars in the Milky Way, given the Galactic core-collapse rate 
\citep{Rozwadowska-2021} as well as recent High-Altitude Water Cherenkov (HAWC)
observations \citep{hawc_catalog}. Their normal radio emission may be beamed away from us, but giant pulses are not from near the polar cap and may have different beaming properties \citep{philippov-2020}. 
The central compact objects of supernova remnants, which 
do not produce normal radio emission, may generate sudden bursts of SGPs
powered by multipolar magnetic fields \citep{cco-snr}. While they show no sign of strong pulsar wind nebulae and appear to have weak dipolar magnetic fields, their strong multipolar magnetic fields offers a potential energy source for bright bursts \citep{vigano-2012}. We emphasize that this scenario is speculative as such bursts have not been observed.
The central compact objects of supernova remnants \citep{cco-snr}, which
do not produce normal radio emission, may generate sudden bursts of SGPs
powered by multipolar magnetic fields. While they do not exhibit strong 
pulsar wind nebulae, 
It is likely that some SGPs from unknown pulsars may be detected by the unprecedented blind search of GReX.
A pulse of $30\rm\, MJy$ in flux corresponds to $\nu
L_\nu = 1.4\times10^{38}\rm\, erg\, s^{-1}$ at a distance of 2 kpc,
so the (non-)detection of such pulses will test whether giant pulse
isotropic-equivalent luminosity can (temporarily) exceed the spin-down
luminosity of the pulsar.

\vspace{0.25cm}
\noindent\textbf{Extragalactic FRBs} GReX's wide, 
shallow survey strategy also allows us to 
probe the nearby, ultra-bright extragalactic FRB population. 
While the primary science function of the instrument is as a 
Galactic explorer, we might expect $\mathcal{O}(1)$ FRBs from 
external galaxies over the instrument's life time. Extrapolating 
from the ASKAP fly's eye survey \citep{shannon18_askap_sample}, 
$N(>\!\mathcal{F})\simeq 5\times10^{3}{\rm\,sky^{-1}\,yr^{-1}} (\mathcal{F}/100{\rm\,Jy\,ms})^{-3/2}$ \citep[see Fig. 3 of][]{lu19_askap_statistics}, to the fluence threshold of GReX, one obtains the detection rate 
\begin{equation}
    N_{\rm det}(>\mathcal{F}_{\rm th}) = 0.45{\rm\,yr^{-1}} \left(50{\rm\, kJy\,ms} \over \mathcal{F}_{\rm th}\right)^{3/2}.
\end{equation}
\noindent This rate is highly sensitive to the logarithmic slope 
of the FRB brightness distribution, as we are extrapolating over nearly 
three orders of magnitude in fluence. If the source counts are flatter than the
Euclidean value of -3/2 in the ultrabright tail, GReX might detect multiple extragalactic FRBs. There may also exist a large 
population of very narrow bursts that ASKAP would have missed 
due to its relatively coarse time and frequency resolution. 

\vspace{0.25cm}
\noindent{\bf Solar Astronomy}
GReX will have continuous coverage of the sun 
with $\sim$\,10\,$\mu$s sampling 
and high frequency resolution over a large bandwidth. 
We will therefore have access to detailed dynamic 
spectra of fast solar phenomena such as 
``millisecond spike bursts'' \citep{ms-spike-1999}. 
We will be able to study Type IV radio bursts, which are likely 
generated through coherent electron cyclotron maser (ECM) emission \citep{ms-spike-1999, solar-typeiv}.

\vspace{0.25cm}

\vspace{0.35cm}

We remark on two questions at the end of this section. STARE2 could see 
approximately 25$\%$ of the northern sky, meaning at most Earth rotational phases 
it would not have seen the FRB-like event from SGR\,1935+2154. This explains the importance of covering the entire sky for rare but bright events. A world-wide GReX network (as described above) can be built for under a million dollars. Brilliant bursts such as FRB\,200428 can be detected via the side-lobes of powerful facilities; indeed, that is how CHIME/FRB discovered FRB\,200428. 
In the northern hemisphere, CHIME/FRB and GReX-I will cover a large portion of the visible sky from 400\,MHz to 2\,GHz, with 100\,MHz of overlap at 750\,MHz. This will allow for future symbioses similar to the joint discovery of FRB\,200428-like events. 

Low frequency facilities such as MWA and LWA naturally enjoy large field-of-view. However, computational costs and interstellar scattering increase as one attempts blind searches at lower frequencies. In the following section we carry out detailed modelling of the 
magnetars whose bursts we hope to detect.
\begin{table}[h]
\caption{GReX: Top-level specifications for search} 
\centering 
\begin{tabular}{l l} 
\\
\hline\hline 
Specification   &  Value\\
\hline
Band            &  0.7--2 GHz \\
Tsys            &  25\,K  \\
Polarization    &  dual/linear \\
field-of-view   &  1.5\,steradian \\
Sampling time   &  32\,$\mu$s (initial) / 8\,$\mu$s (hardware)  \\
Channel width (700-1650\,MHz)   &  116\,kHz \\
Channel width (1650-2000\,MHz)  &  42\,kHz \\
Fluence         &  100 kJy for 1-ms burst \\
Timing          &  link to GPS ($\pm$\,10\,ns)\\
\hline
\end{tabular}
\end{table}

\section{Galactic Radio Explorer: Implementation}

\label{sect-hardware}
The basic unit of GReX has a field-of-view of $\sim$\,1.5 steradian and a frequency range of 0.7--2\,GHz. We will search data down to 
$\sim$\,10\,$\mu$s. RFI rejection and crude localization (via
timing) requires three units separated by at least one hundred kilometers.
We call such a triplet as ``cluster". To cover the entire sky would
require eight clusters (four in the North separated by 80 degrees
in longitude and four in the South). Thus, a full-up GReX network would
have 24 unit systems. Reducing the unit cost is important 
and is an engineering requirements in the GReX pilot phase.
During GReX Phase I we will build a cluster with a 3:1
radio band and, thanks to novel LNAs, sensitivity at least twice
that of STARE2. Combined, the increase in sensitivity relative
to STARE2 could be as much as a factor of five. We are able 
to achieve such improvements by piggy-backing on the 
advances in electronics and antenna design 
provided by DSA-110 and DSA-2000. 
Leveraging this work, we will have the lowest system noise temperature (approximately 20\,K) 
yet achieved without cryogenic cooling in the proposed frequency range.
While this is our starting point, 
there is enough flexibility to cater the feed and back-end 
design to GReX's science goals, for example 
trading field of view for forward gain or shifting the frequency range on a 
per-site basis to avoid RFI. Similarly, the RFI excision 
for each cluster of GReX antennas will likely need to 
be tailored to the local RFI environment, even if the algorithms 
are common across locations.

We have chosen the GReX configuration 
over, for example, a focal-plane phased-array without a reflector\footnote{https://old.astron.nl/r-d-laboratory/ska/embrace/embrace} 
for the following reasons: The added complexity and cost related to developing such a system would be non-trivial, as would the increased 
cost of digitization, channelization, and beamforming compute hardware that follow from having more feeds. The current feed design 
provides a frequency-independent beamwidth, and a single Stokes I 
beam requires only two digital backends per antenna, 
one for each polarization. Therefore, 
if our goal is to monitor the whole sky continuously with 
high time and frequency resolution, we find that the most 
effective way is to deploy simple single radiometer systems 
around the world. Another suggestion for an 
all-sky Galactic FRB survey relied on Citizens-Science and 
cellular communication devices to search for $\sim$\,10\,GJy 
bursts \citep{maoz-2017}. However, the detection of ST\,200428A just above the 
STARE2 threshold established that the Galactic FRB brightness 
distribution cannot be very flat and Giga bursts are likely 
exceedingly rare.

With a world-wide GReX system that includes both more sky coverage and 
more exposure to the Galactic plane, we anticipate more than 
an order of magnitude increase in detection rate over STARE2. 
To achieve this, we aim to create an assembly kit that 
can be shipped at cost to interested parties around the world. 

\subsubsection{Wideband Antenna}
\begin{figure}[htbp]
 \centering
  \includegraphics[width=0.8\textwidth]{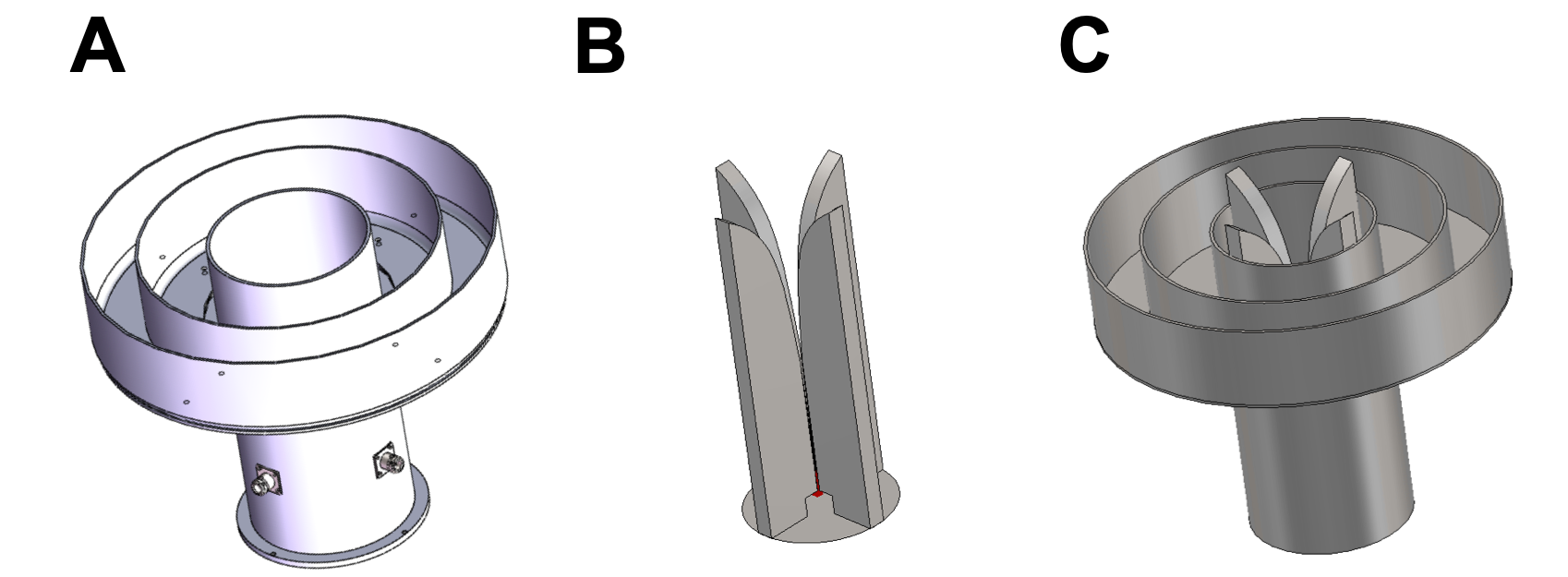}
   \caption{\small ({\it A}) The current STARE2 low-cost ``cake-pan''
   antenna which delivers uniform beam-width and excellent efficiency
   but over a limited bandwidth. ({\it B}) Quad-ridge structure
   added to the cake-pan delivers wider bandwidth while
   maintaining a low-cost design. ({\it C}) The assembled quad-ridge
   choke horn structure concept to be used for GReX.}
 \label{fig:feed}
\end{figure}

The current ``cake-pan'' antenna of STARE2 is fabricated from a
6$^{\prime\prime}$-diameter aluminum pipe surrounded by two cake
pans which reduce spillover and increase efficiency, providing a
low-cost solution that delivers uniform beam-width across the
256\,MHz band. 

\begin{figure}[htbp] 
 \centering
  \includegraphics[width=2.5in]{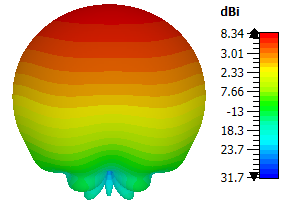}
     \qquad
  \includegraphics[width=3.5in]{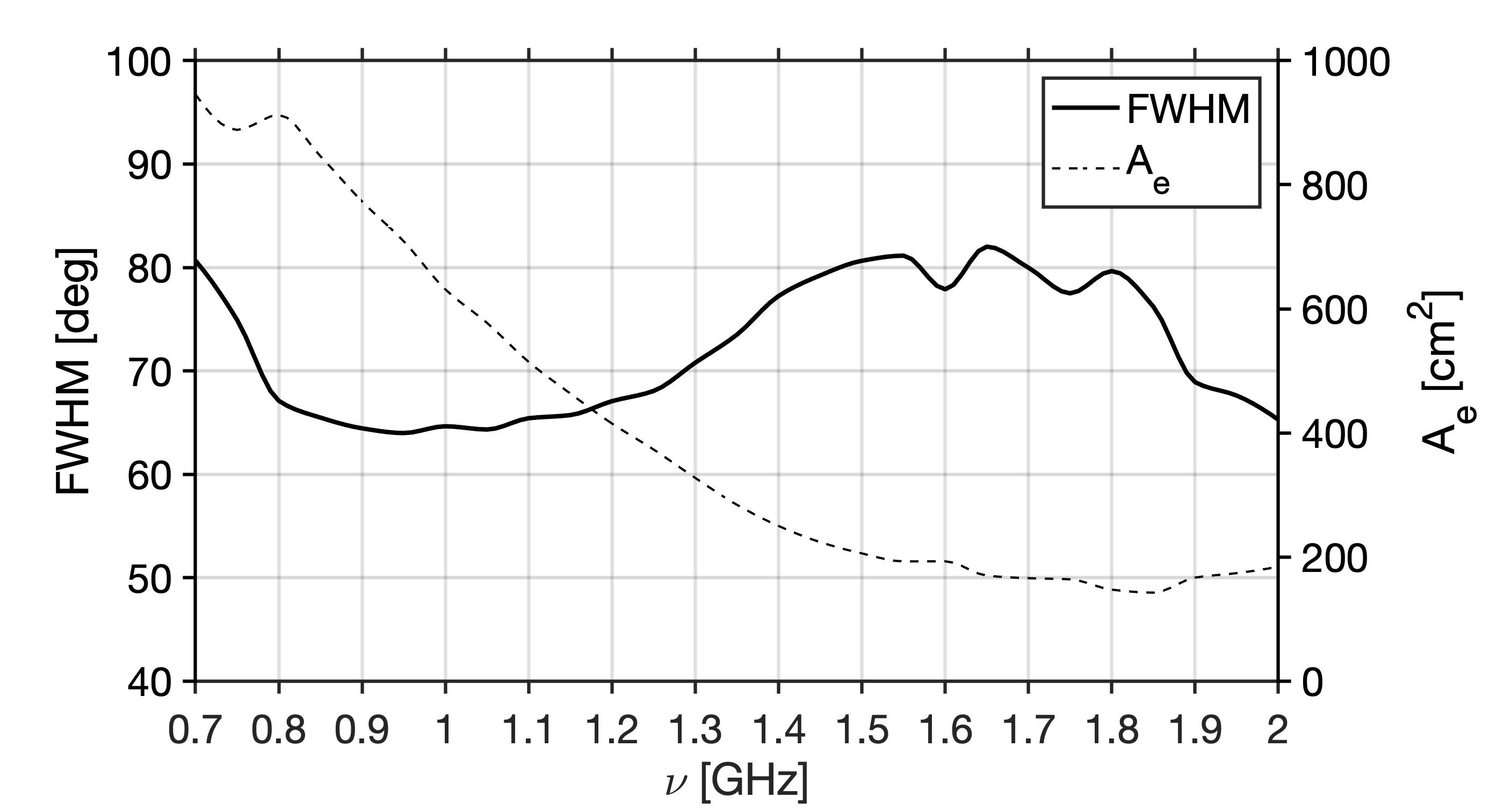}
   \caption{\small (left): Polar representation of the beam at
   1.4\,GHz. right): Simulation of beam full-width-at-half maximum (FWHM)
   in degrees and effective area, $A_e$, as a function of frequency,
   $\nu$ (in GHz).}
 \label{fig:BeamPattern} 
\end{figure}

Here, we propose to use a quad-ridge horn with a choke-ring structure (Figure~\ref{fig:feed}) for wide-beam performance over a wide frequency range. The design is based on the quad-ridge flared horn (QRFH) technology, developed at Caltech by graduate student Ahmed Akgiray\footnote{Akgriay's 2013 Caltech PhD is a convenient reference: \url{https://thesis.library.caltech.edu/7644/}} and his advisor Dr.\ Sandy Weinreb. To take advantage of the cost-effective design of the cake-pan antenna, the quad-ridge structure will be integrated with the choke-rings (Figure~\ref{fig:feed}). The choke-ring structure reduces side and back-lobes resulting in a near-symmetric beam pattern. In Figure~\ref{fig:BeamPattern}, the beam is exemplified at 1.4\,GHz, and the wide, near-constant, full-width-at-half maximum (FWHM) presented over frequency. The quad-ridge structure enable dual linear polarization within a compact footprint, and good impedance match to low-cost 50~$\Omega$ coaxial connectors resulting in low input-reflection coefficient over the wide frequency band. To reduce the contribution of noise from ground pick-up to only a few K (Figure~\ref{fig:ground_pickup}), a shield underneath the antenna will be used.

\subsubsection{Low Noise Amplifiers (LNAs)}
A break-through in low noise amplifier (LNA) technology has occurred in the past few years and  enables a sensitivity improvement of a factor of 3 compared to the monitors used for the previous STARE2 detection.  This can be achieved by reducing the noise temperature of the LNA from 32\,K to 10\,K and contributions of feed spillover and other losses from 28\,K to 10\,K to improve the  system noise temperature from 60K to 20K.  This can be accomplished without the utilization of cryogenic coolers which are costly, require much AC power, and require considerable maintenance. 

\begin{figure}[htbp]
\centering
 \includegraphics[width=0.65\textwidth]{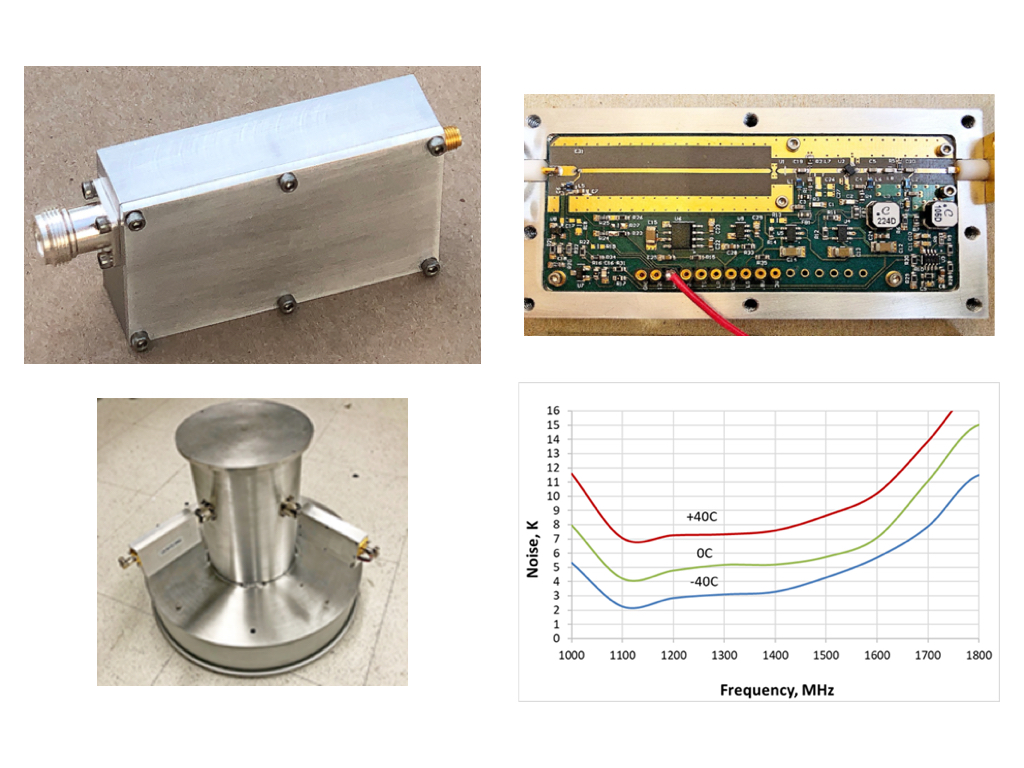}
   \caption{\small The DSA-110 LNA showing exterior view (top left), interior view (top right), as mounted on the DSA-110 feed (bottom left), and noise temperature vs frequency at 3 temperatures (bottom right). The LNA requires no wires for bias or control; the bias of +5V is supplied on the output coaxial line with control of an internal noise calibration source by a 32kHz  tone on this output line.}
 \label{fig:LNAs}
\end{figure}

This LNA break-through has been demonstrated in the DSA110 array at the Caltech Owens Valley Radio Observatory (OVRO) where a system noise of 25\,K has been measured on 25 dual-linear polarization 4.6 m paraboloidal reflector antennas operating in the 1.28 to 1.53 GHz frequency range.  The LNA for this project is summarized in Figure~\ref{fig:LNAs} and fully described in a paper \citep{weinreb-2021}. The key elements for this low noise are an  extremely high performance (Fmax of 550 
GHz) high-electron-mobility-transistor (HEMT) on an indium-phosphide (InP) substrate Type pH-100 discrete InP HEMT
\footnote{Diramics AG, Zurich, https://diramics.com/products/} and an extremely low loss ($<$0.05\,dB)  matching network from the chip transistor pads to the input connector, A brief description of the LNA has been presented at a workshop \citet{weinreb-2021}\footnote{https://events.mpifr-bonn.mpg.de/indico/event/154/session/4/contribution/27}.

\begin{figure}[htbp] 
\centering
 \includegraphics[width=0.75\textwidth]{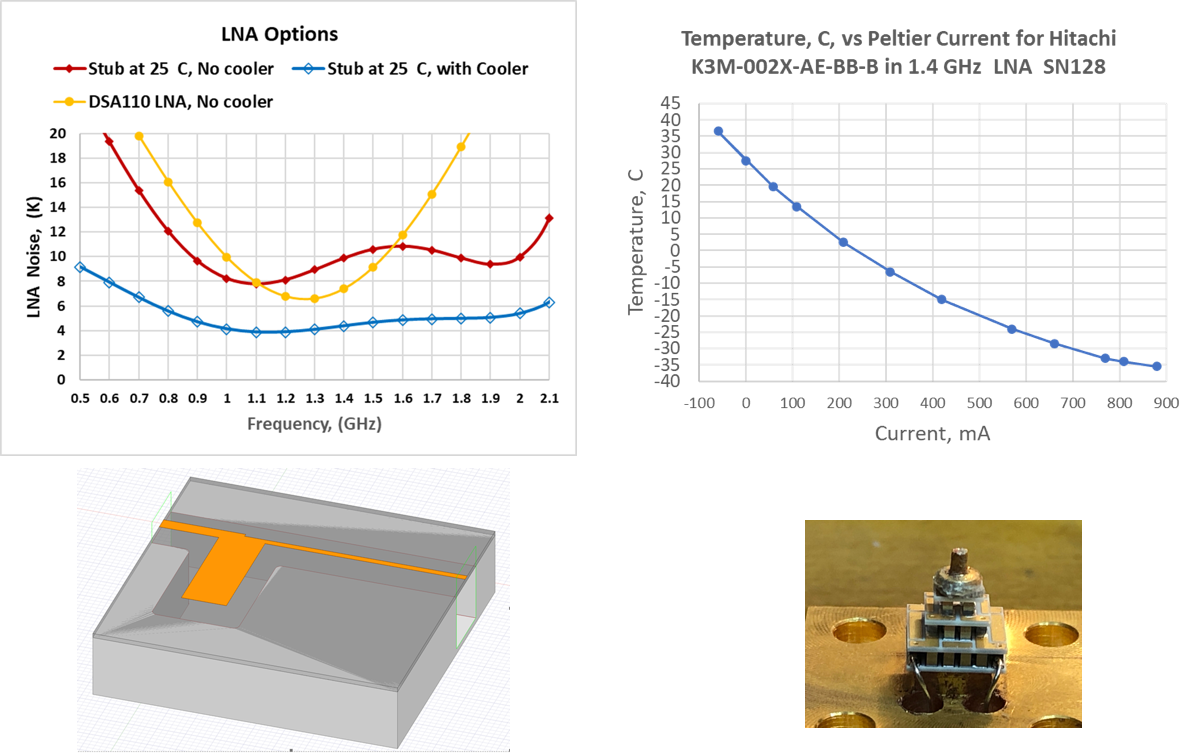}
   \caption{\small (left) Modeled LNA noise temperature in red without a microcooler, averaging 10.1 K, and with the shunt stub input matching network shown below. (right) Measured physical temperature of a transistor chip mounted on the 6mm high microcooler shown below.}
 \label{fig:widebandLNAs}
\end{figure}

\begin{figure}[htbp]
 \centering
 \includegraphics[width=0.6\textwidth]{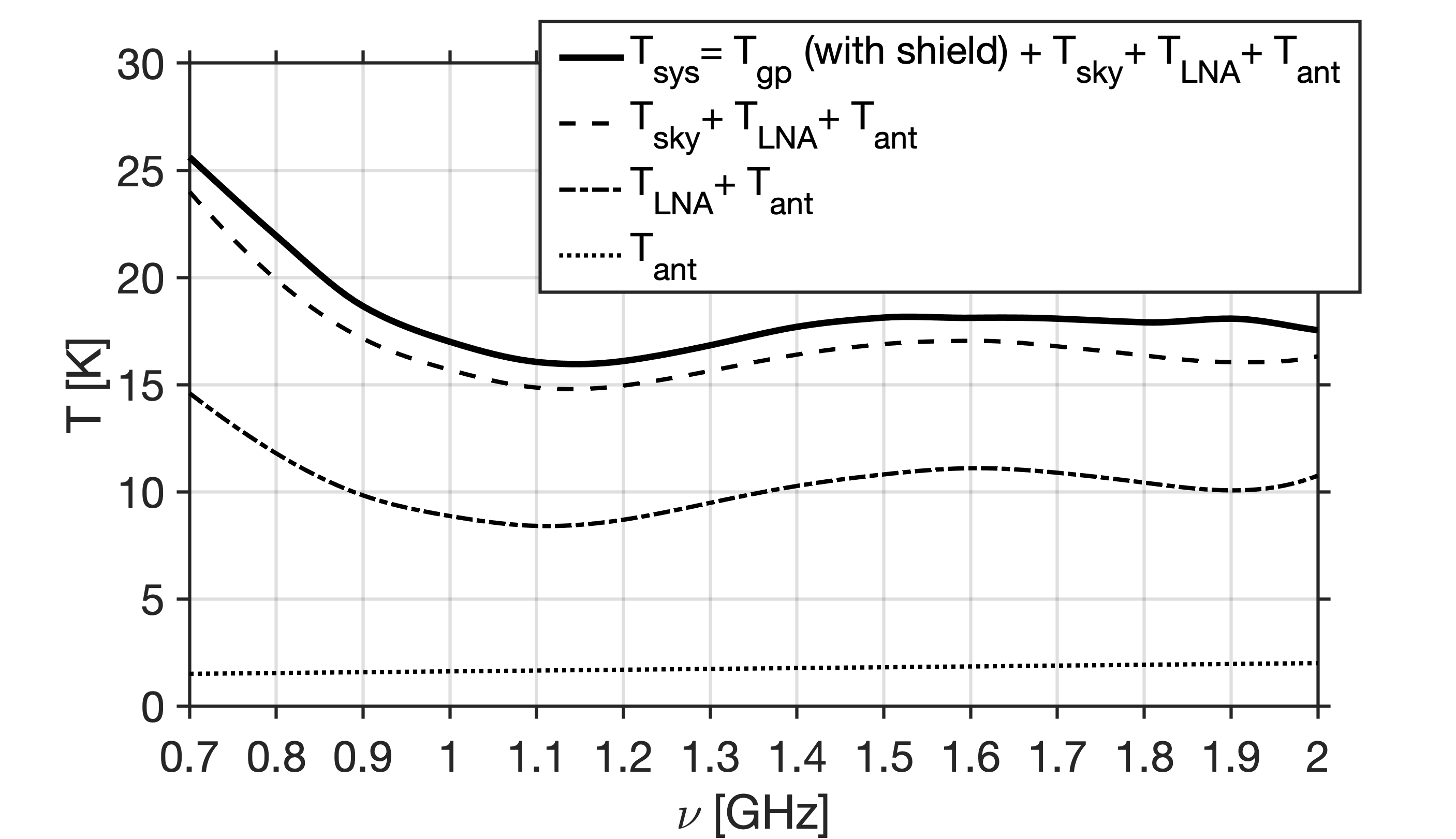}
  \caption{\small Simulation of system temperature, separated as contributions from ground pick-up ($\rm T_{gp}$), sky ($\rm T_{sky}$; i.e. CMB, Galaxy, atmosphere), the low-noise amplifiers ($\rm T_{LNA}$), and losses in the wideband antenna ($\rm T_{ant}$). A ground radiation shield in the form of a square with 4~m sides and 0.3~m, slightly angled, walls reduce the ground pick-up to a few K. A study for optimal shield-material and shape will be done to weight costs (mesh or solid material; curved or flat) and benefit (${\rm T_{gp}}$).}
 \label{fig:ground_pickup}
\end{figure}

Our goal for GReX is an LNA for 0.7 to 2\,GHz with frequency-averaged noise temperature under 10K. This is challenging due to the required bandwidth of the input matching network to transform 50\,ohm impedance of the feed to the optimum impedance driving the transistor which is known from previous studies. This network must have extremely low loss with the realization that 0.1\,dB of loss adds 7\,K to the noise temperature. There are two approaches to this challenge illustrated in Figure~\ref{fig:widebandLNAs}: 1) a more complex, but very low loss, input matching network, and 2) cooling just the transistor chip to -40\,\textdegree C using a Peltier effect solid-state micro-cooler. Using computer-aided-design optimization with high-frequency structure simulator (HFSS) electromagnetic modeling software, a shunt stub Tee input network has been optimized to give a frequency-averaged noise temperature of 10.1K with operation at 25\,\textdegree C or approximately 5 K less noise by cooling to -40\,\textdegree C. Analytic models of the loss and capacitance of the Tee connection were not available so a field analysis utilizing finite elements to solve the Maxwell equations was required. A major challenge with cooling to -40\,\textdegree C is the condensation of water and ice on the transistor. To prevent this over a long period of time, both high vacuum or pressurization with a low thermal conductivity gas such argon or xenon are being investigated. There is risk to the cooling and our next step will be to construct a prototype LNA with the wideband input network, measure the LNA noise in a laboratory test setup, and then, measure the system noise with the feed including the ground radiation shield. 

Following the LNAs additional amplification, filtering, and perhaps frequency conversion will be required to drive the A/D converters and the digital spectrometer. These can be more conventional RF design modules similar to those used in DSA-110 but with wider bandwidth. 

\subsubsection{Digital back-end}

We propose to use SNAP-2 platform\footnote{This includes a Kintex
Ultrascale FPGA fed by dual FMC-mounted 5GAD ADCs. Both are supplied
by the Institute for Automation in Beijing, China } which allows
direct sampling (5 Giga-samples per second, or 5\,Gsps) of the
entire 0.2--2\,GHz RF signal to process the 0.7--2 GHz bandwidth.
The ADCs sample at 10-bit precision which we believe
is sufficient to excise strong RFI that is present in the band of
interest to us. We will implement
a standard four-tap polyphase filter-bank, using a channelization scheme described below.
Following detection in each
polarization we will integrate to the desired time resolution, optimally
re-quantized to 8-bit precision, and streamed across a 40 GbE direct
connection to a computer server (data rate of 8 Gbps). The use of
a 40 GbE connection permits streaming voltage
data to the server for buffering for the entire usable bandwidth.

As we hope to search the radio sky at $\mathcal{O}(10\,\mu s)$, we must account 
for intrachannel dispersion smearing. This effect refers to temporal broadening 
caused by the dispersion delay between adjacent frequency channels and is given by the following 
equation,

\begin{equation}
    t_{\mathrm{DM}} = 8.3\, \mathrm{DM} \left( \frac{\Delta\nu}{1\,\mathrm{MHz}} \right )
    \left( \frac{\nu_c}{1\,\mathrm{GHz}} \right )^{-3} \, \, \mu \mathrm{s}.
\end{equation}

\noindent where $\nu_c$ is the observing frequency and $\Delta\nu$ is the frequency 
resolution. The total smearing is then the quadrature sum of the sampling time and the 
intrachannel dispersion term,

\begin{equation}
    t_{\mathrm{smear}} = \sqrt{t^2_{\mathrm{samp}} + t_{\mathrm{DM}}^2}.
\end{equation}

\begin{figure}[htbp] 
\centering
 \includegraphics[width=0.6\textwidth]{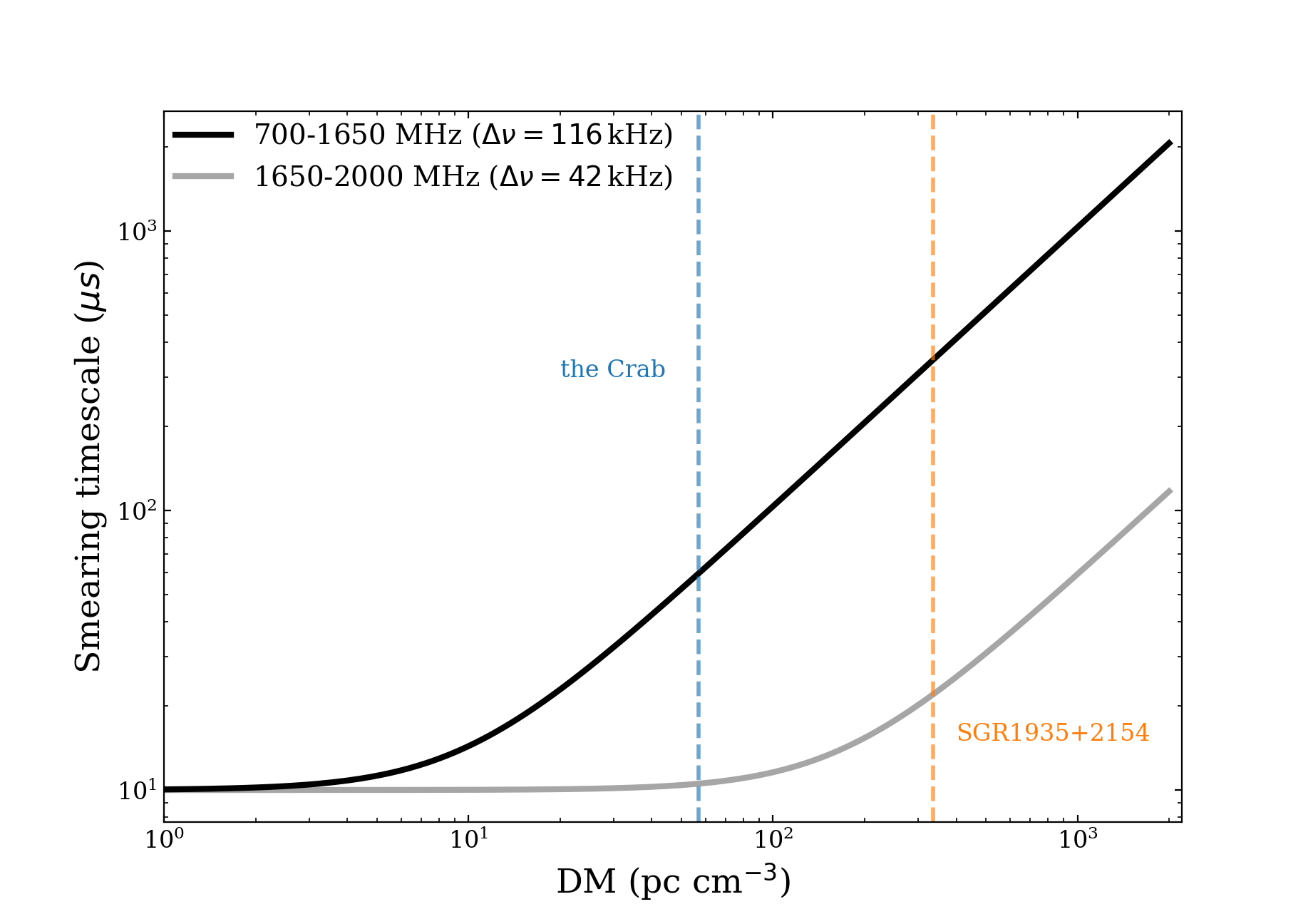}
   \caption{Temporal smearing curves for the two sub-bands of GReX plotted as 
   a function of DM. The 
   top 350\,MHz of our band will have finer spectral channels, preserving our 
   ability to search down to tens of microseconds even for high-DM sources.}
 \label{fig:smearing}
\end{figure}

\noindent With 16,384 channels across the full band centered on 1.35\,GHz, 
this term is 100\,$\mu$s at the DM of SGR\,1935+2154, and we could not 
achieve our proposed $\mathcal{O}(10\,\mu s)$ timescale at such a DM. In order to preserve our 
high-time resolution search for sources in the Galactic plane that may also have high DM, we will 
implement a hybrid channelization scheme. The top 350\,MHz of our band 
will have 42\,kHz resolution (8192 channels) and the bottom 950\,MHz will have 
116\,kHz resolution (8192 channels as well). The single pulse search will combine candidate 
information between the bands, without loss of information. 
The smearing curves for both sub-bands 
are plotted as a function of DM in Fig.~\ref{fig:smearing}.


\subsubsection{Algorithm \& Analysis Computer}
 
GReX’s real-time detection pipeline and computing architecture will
be tailored to the instrument’s broad radio band and high time
resolution; the traditional
\texttt{Heimdall} package,
which was used for STARE2, is not well suited to the wide band
of GReX, the top of whose frequency range is three times 
the lower frequency. We will employ a ``sub-band’’ dedispersion algorithm, in
which GReX’s large frequency range will be apportioned into uniform
chunks in $\nu^{-2}$ that will be searched independently and later
combined. To this end, we will customize the fast dedispersion
algorithm \texttt{FDMT} \citep{zackay-fdmt}, which
allows for sub-band searching and optimally detecting FRBs with
significant frequency structure using the Kalman
filter\footnote{https://bitbucket.org/bzackay/kalman\_detector/src/master/}.
GReX will thereby serve as a proof-of-concept for future ultra-wideband surveys such
as DSA-2000 and the funded Canadian project CHORD \citep{vlg+19}.

The computer server will be equipped with sufficient processing
power to reject impulsive RFI in real time, and to search for FRBs.
Data will be piped through the analysis software using the \texttt{psrdada}
framework. The RFI rejection, dedispersion, and pulse finding will
be implemented on an Nvidia Quadro RTX4000 GPU. 
The hardware is well suited to 
a final-stage machine
learning classifier which will send out reliable, real-time triggers
to other facilities \cite{connor-ML-2018}. RFI excision will 
be done using a common set of modular algorithms. Since we expect 
each GReX site to have different RFI characteristics, the exact parameters 
of those algorithms will be tailored to the site. While the final set of 
sites is not yet known with certainty, we plan to choose locations whose RFI environment 
is relatively clean.

We plan to reserve roughly a dozen DM channels to 
be coherently dedispersed when known pulsars are in the beam, 
enabling us to search for super-giant radio pulses without 
the deleterious effects of instrumental smearing. The DMs will 
correspond to known millisecond pulsars and young pulsars, 
the sources most likely to emit giant pulses. 

Our design allows
for upgrades to be explored by the group, in response to the evolving
scientific landscape. For example, we can modify the digital firmware
to stream voltage data for buffering on the server, such that when
a pulse is detected the data can be analyzed with better time and
frequency resolution. Additionally, we have sufficient processing
power on the FPGA to derive full-polarization data, if needed.
\vspace{1cm}
\subsection{Future Stations}
 \label{sec:FutureStations}
 
One of the goals
of GReX is to expand University-based radio observatories in the 
United States. 
Specifically, we envisage a future effort to place
unit systems at sites that are managed or accessible to Universities,
with emphasis on state universities in the US with the intention 
of accelerating radio astronomy education for undergraduates. We also 
feel that a GReX cluster in Canada, for example at the Algonquin Radio Observatory (ARO) would help with both sky coverage and would allow for 
continuous coverage at 400--2000\,MHz in the Northern Hemisphere.
RFI studies will be undertaken of potential future site. GReX will serve this purpose explicitly as a test antenna at the proposed DSA-2000 site, Big Smoky Valley, Nevada, where it will both act as an RFI monitor and carry out its primary science. For clusters within $\sim$\,thousands of km of one another, we will offset the antennas in pointing to minimize overlap on sky.

During GReX Phase II we plan expand beyond the United States, 
starting with Australia and India in order to build up 24/7
coverage of the Galactic plane, where the majority of magnetars 
reside. Clusters of antennas in both Western Australia and Tasmania 
would provide 120\,$\deg$ of coverage in right ascension, if they 
each point 7.5\,$\deg$ off-zenith. A set of three 
antennas in India could be located at GMRT. We also hope to deploy GReX 
instrumentation in Western Europe, the Middle East, and elsewhere until 
we have an international network spanning the full sky's 4$\pi$ steradians. We anticipate that this will require roughly 25 
hardware kits.

\begin{figure}[htbp] 
 \centering
  \includegraphics[width=0.65\textwidth]{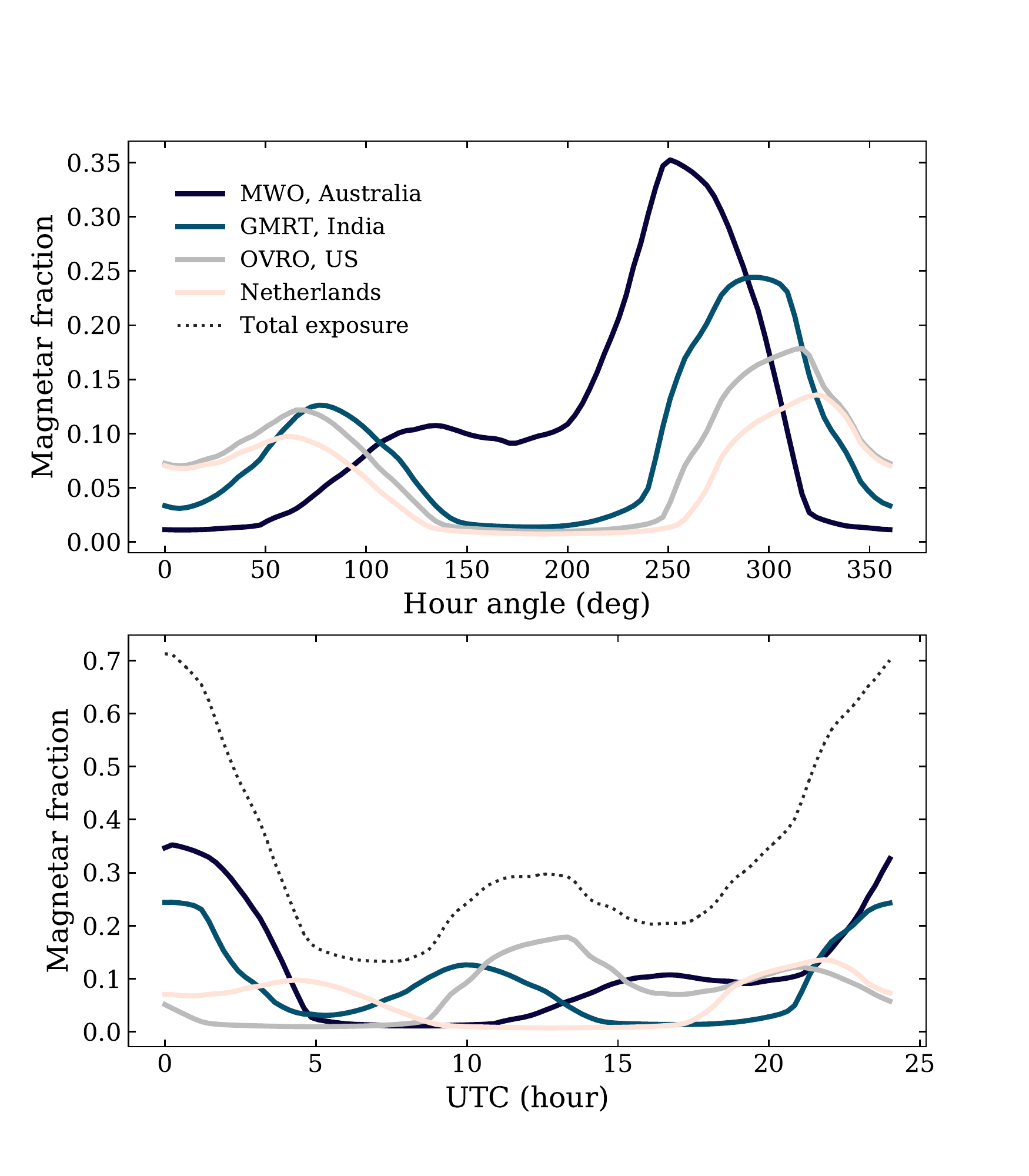}
   \caption{\small Modelling of the fraction of Galactic magnetars 
   that are within the FoV of a GReX antenna for different potential 
   cluster locations around the world. The top panel shows 
   this fraction as a function of hour angle (HA). The bottom 
   panel shows the same fraction vs. UTC, accounting for 
   the different RAs to which each location is exposed 
   at a given time. The dotted line is the combined visible fraction 
   of Galactic magnetars for the whole GReX-II array.}
 \label{fig:magnetar-exposure}
\end{figure}
\section{Modelling}
\label{sect-forecast}
By sending GReX clusters to southern latitudes we will gain 
exposure to the bulk of the Galactic plane, where magnetars 
reside. We also aim to increase our coverage 
in right ascension by deploying clusters at a range of 
longitudes. Using molecular gas and star formation as a 
proxy for the magnetar distribution in the Milky Way, 
we model the fraction of Galactic magnetars to which 
GReX will be exposed at a given time. Combining this with 
the increased sensitivity of each unit, we can  
forecast the expected Galactic FRB rate and speculate on 
the GRPs we can find from pulsars.

We take the Planck HFI 857\,GHz map and assume we have 
clusters in the Southwestern United States,  
India, Western Europe and Australia. Based on Figure~\ref{fig:BeamPattern}, we take the GReX antenna's FWHM to be 80\,deg. We compute 
to total visible material in the Planck 857\,GHz map at a 
given time to estimate the fraction of Galactic magnetars 
that are observable at each cluster. The top panel of 
Figure~\ref{fig:magnetar-exposure} shows this fraction 
as a function of hour angle at four locations. 
Clearly, the heightened exposure to the inner 
Galactic plane gives the southern stations an advantage. 
The bottom panel show the same fraction, but as a function 
of coordinated universal time (UTC), including the 
instantaneous total exposure of all stations combined. For more than half 
of the time, this configuration is exposed to more than 30$\%$ of 
the magnetars in our Galaxy. Eventually the network will see the whole sky all of the time.

\subsection{Event rates}
The system temperature of GReX is expected to 
be $\sim$\,2.6 times lower than that of STARE2, thanks 
to the aforementioned improvements in front-end electronics. 
Its antennas will have five times as large a frequency band, and the 
addition of southern clusters will give GReX more instantaneous sky coverage and increased exposure to 
the Galactic plane. Extrapolating from the STARE2 detection of 
one burst in 448 days on sky, we can estimate the detection rate of GReX-I and GReX-II. 
\begin{figure}[htbp]
 \centering
      \includegraphics[width=0.75\textwidth]{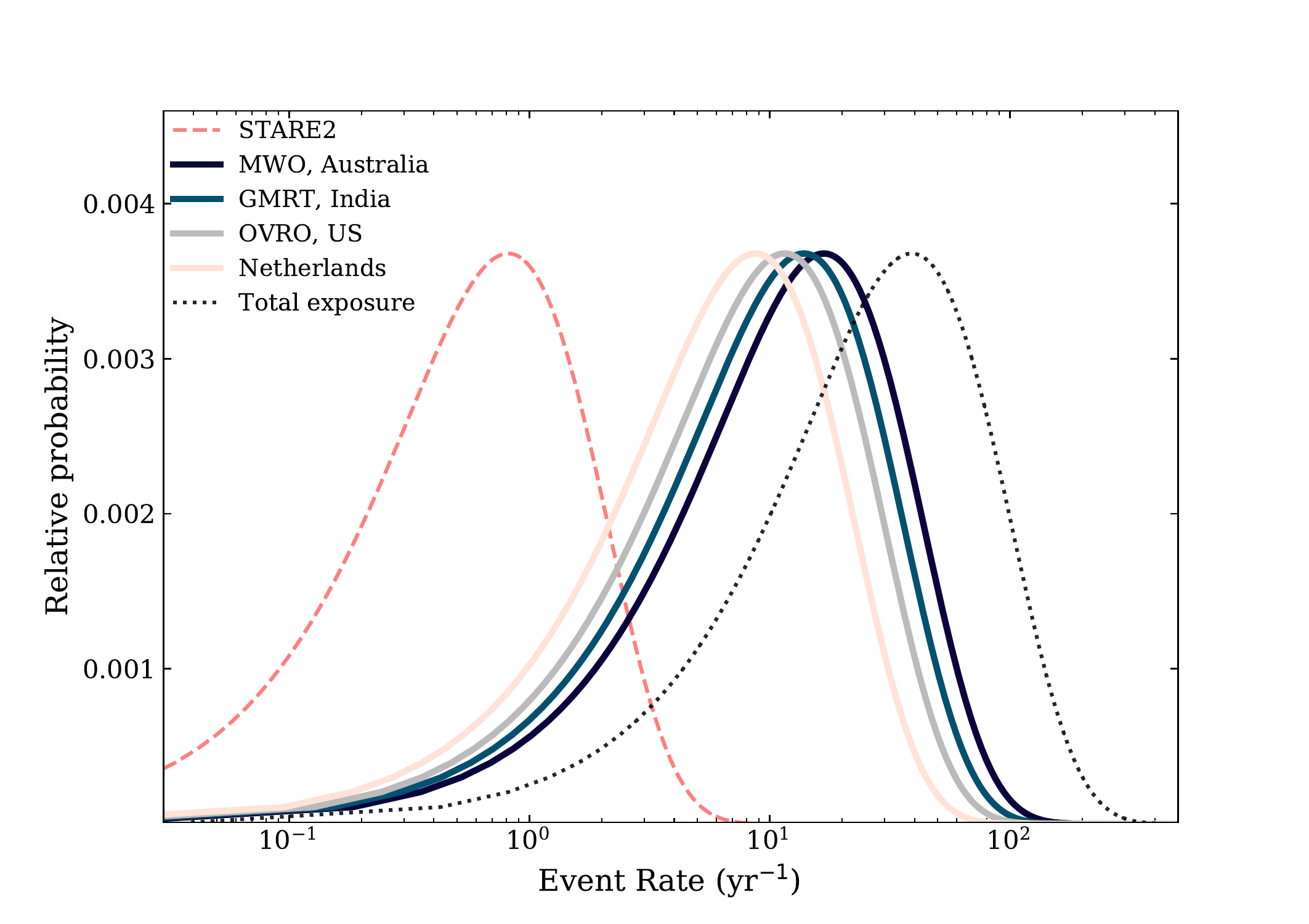}
  \caption{\small The event rate distributions 
            based on one FRB-like event detected 
            by STARE2 in 448 days on sky. The relative 
            probabilities vs. detection rate are plotted for 
            individual GReX clusters around the world (solid) 
            as well as a hypothetical GReX-II array (dotted) 
            that has antennas in Australia, the Netherlands, 
            India, and the United States. Even if 
            FRB\,200428 were a rare event and STARE2 and 
            CHIME/FRB were ``lucky'' to have detected it, 
            the full array will likely find multiple such 
            events per year.}
 \label{fig-rates}
\end{figure}
The detection rate is given by the product of the survey’s field of view, $\Omega$, and the source density on sky above the instrument’s detection threshold, $N(>s_{\rm min})$. We take the source brightness distribution 
to be a power-law such that, $N(>s_{\rm min})\propto s_{\rm min}^{-\alpha}$, where $s_{\rm min}$ is given by the radiometer equation.
Assuming a baseline rate for STARE2, $\mathcal{R}_{\rm S2}$, the 
detection rate of GReX-II will be a sum over all antenna clusters around the world, weighted by their average exposure to the Galaxy's magnetars 
compared to STARE2. We call this weight for the $i^{\rm th}$ GReX cluster location, $w_i$. This gives,

\begin{equation}
\mathcal{R}_{\rm GReX} = \mathcal{R}_{\rm S2} \sum^{n_{clust}}_{i}\,w_i\,\frac{\Omega_i}{\Omega_{\rm S2}} \left ( \frac{\mathrm{SEFD}_{S2}}{\mathrm{SEFD}{i}}\sqrt{\frac{\mathrm{B}_i}{\mathrm{B}_{S2}}} \right )^{\alpha},
\label{eq-rate}
\end{equation}

\noindent where SEFD refers to the system-equivalent flux density, or the ratio of system temperature to gain, T$_{\rm sys} / G$. Assuming each GReX unit has the same FoV as STARE2 and 
that the pointings are mostly independent, we get

\begin{equation}
\mathcal{R}_{\rm GReX} = 1/448\,\mathrm{days}^{-1}\left ( 2.6\sqrt{5} \right )^{\alpha} \sum_{i}\,w_i\,.
\end{equation}

\noindent We note that the $\sqrt{5}$ factor implicitly assumes that the burst spectrum is both flat and broadband, which may be optimistic. Still, we do not have a good model for the spectral structure of FRBs in general, and ST\,200428A/FRB\,200428 was brighter at 1.4\,GHz than at 0.6\,GHz, so for the sake of simplicity we have assumed equal power across the GReX band. Equation~\ref{eq-rate} computes the maximum-likelihood value 
of the event rate based on STARE2's sole detection, but we must include the uncertainty associated with 
just one burst. Using Bayes' theorem we know,

\begin{equation}
    P(\mathcal{R} | N) = \frac{P(N | \mathcal{R}) P(\mathcal{R})}{P(N)},
\label{eq-poisson}
\end{equation}

\noindent where in this case $N=1$. Taking a flat prior on 
$\mathcal{R}$ and assuming the detection 
of new bursts follow Poissonian statistics, we can 
invert Equation~\ref{eq-poisson} to get,

\begin{equation}
    P(N\!=\!1\,| \,\mathcal{R}) = \mathcal{R}\,e^{-\mathcal{R}}.
\end{equation}

\noindent By scaling $\mathcal{R}$ for GReX using 
Equation~\ref{eq-rate}, we can calculate the probability 
distribution in detection rate after the improved sensitivity 
and exposure to Galactic magnetars.
This is shown in Figure~\ref{fig-rates}.

The forecasting we have presented has extrapolated from STARE2's single detection 
of ST\,200428A/FRB\,200428 using the modelled distribution of magnetars in 
the Milky Way. The result is a Poissonian confidence interval for GReX, 
but this range is \textit{not} the rate of unique GReX detections. A burst 
in the Northern Hemisphere could also be detected by CHIME/FRB, barring observing 
frequency differences. Furthermore, an X-ray all-sky monitor might observe high energy 
activity from a Galactic magnetar, allowing other radio instruments to spend time on 
that source during its active phase and detect a radio burst that GReX would also see.
While this detracts from our unique discovery capability, it is likely that multiple detections of the same event will add scientifically to the discovery. Together, CHIME/FRB and GReX will span the full range between 400 and 2000\,MHz, overlapping only between 700--800\,MHz.

\section{Summary}
We have proposed GReX, a radio all-sky monitor that will detect the 
brightest bursts in the Galactic sky on sub-millisecond timescales. 
GReX Phase II will be an international network of ultra-wideband, high performance
antennas that will continuously search for Galactic FRBs and super giant pulses 
from radio pulsars. Each hardware component will be low-cost and replicable, 
such that clusters of at least three 
GReX antennas can be shipped as 
an assembly kit around the world to span a wide range of latitude and 
longitude. The software and firmware back-end will also be standardized such that 
each system will require little intervention once running. We expect to find multiple new FRB-like events from Galactic magnetars each year. As the first 
wide-field, blind single-pulse survey at the microsecond level, we should to find new super giant pulses 
from Galactic pulsars, as well as previously-unknown 
phenomena. 

\section*{Acknowledgements}
We thank Dale Gary and Dan Werthimer for helpful discussions, 
as well as an anonymous referee for valuable advice on the manuscript.

\bibliography{grex}{}
\bibliographystyle{aasjournal}



\end{document}